\documentclass[preprint,nofootinbib,APS]{revtex4}
\usepackage{amsmath,amssymb}

\def\wpow#1#2{{#1}^{\wedge #2}}

\begin{document}

\title{
Generalized Instantons on Complex Projective Spaces
}

\author{Hironobu Kihara}
\affiliation{ Korea Institute for Advanced Study\\
207-43 Cheongnyangni 2-dong, Dongdaemun-gu, Seoul 130-722, Republic of Korea
}
\author{Muneto Nitta}
\affiliation{Department of Physics, Keio University, Hiyoshi, Yokohama, Kanagawa
 223-8521, Japan}

\preprint{KIAS-P08045}
\date{July 08, 2008}

\begin{abstract}
We study a class of generalized self-duality relations 
in gauge theories on the complex projective space 
with the Fubini-Study metric. 
Our theories consist of only gauge fields with gauge group ${\rm U}(n)$. 
The pseudo-energies which we consider 
contain higher orders of field strength and 
are labeled by an integer $p$ smaller than or equal to $[n/2]$. 
For making the Bogomol'nyi completion 
we need non-single trace terms in the pseudo-energies, 
unlike the models defined on spheres, which were studied previously.
We construct an explicit solution of codimension $2n$ 
to generalized self-duality equations as Bogomol'nyi equations, 
by using a part of the spin connection. 
\end{abstract}

\maketitle

\section{Introduction}

The Yang-Mills instantons \cite{Belavin:1975fg} play 
important roles in field theory and string theory \cite{Dorey:2002ik}. 
The solutions satisfy the (anti)self-duality relation
$F = \pm *F$ in four-dimensional Euclidean space, 
which is a set of non-linear first order partial differential equations. 
Generalizations of the duality relation 
in higher dimensional spaces 
have been discussed in several ways. 
One way is to find an antisymmetric invariant tensor $T$ of rank 4 
on the base space, 
and to consider the first order relations $F = \pm T F$ 
where the Hodge duality is not used in general \cite{Corrigan:1982th}. 
Another way introduced by Tchrakian 
\cite{Tchrakian:1978sf}--\cite{Radu:2007az} 
is to consider Hodge dual relations 
(generalized self-duality relations) between 
higher order terms of the field strength $F$.  
Although these are not of first order anymore,
an advantage is that no extra tensor like $T$ is needed.
An explicit example of the octonionic instantons 
satisfying generalized (anti-)self-duality relations $F^2 = \pm * F^2$ 
was obtained on eight dimensional sphere $S^8$ \cite{Grossman:1984pi}. 
Recently we have studied 
a monopole-like solution of codimension five in 
generalized self-duality relations  
$* F^2 = \pm D_A \phi$ (with $\phi$ Higgs scalars)
as Bogomol'nyi equations in gauge systems 
with higher derivative couplings \cite{Kihara:2004yz}. 
We have also constructed an explicit solution of 
an ``instanton" of codimension six, satisfying 
generalized self-duality relations 
$F^2 =\pm \gamma_7 * F$ 
with SO(6) gauge group on $S^6$ \cite{Kihara:2007di} 
and have applied it to a compactification 
$M_4 \times S^6$ or $AdS_4 \times S^6$ \cite{Kihara:2007vz}. 

In this paper 
we study the Tchrakian's type self-duality relations 
of ${\rm U}(n)$
 gauge theory on general manifolds 
of dimension $2n$, 
and construct an explicit solution 
of codimension $2n$ 
on the complex projective space ${\mathbb C} P^n$.  
Since Hermitian matrices are closed in their multiplications, 
higher powers of the field strength $F$ 
are elements of Lie algebra ,  
in the case of ${\rm U}(n)$ 
 gauge group.
The previous example of SO(6) is a special case because of the local isomorphism; SO(6)$\simeq_{\rm local}$ ${\rm SU}(4)$. 
We consider pseudo-energies which consist of single and double trace terms 
of field strength $F$ 
although one could consider triple or higher trace terms in general. 
In previously known examples 
only single trace term was considered. 
We see that 
double-trace term is needed in the pseudo-energy
in order to make the Bogomol'nyi completion 
and it is enough to construct a non-trivial topological solution.
In this kind of theories, self-duality relation is given as a relation between the Hodge dual of one power of the field strength and another power of it 
where we may need modification by subtracting trace parts.   
Such a self-duality relation has been already discussed 
in \cite{Bais:1985ns,Ma:1990ja}. 
We revise them and find that our solution has non-vanishing 
pseudo-energy. 

Our previous solution on $S^6$ was applied to 
compactification $AdS_4 \times S^6$ \cite{Kihara:2007vz} 
in the spirit of Cremmer and Scherk \cite{Cremmer:1976ir}.
There exists a certain relation between a coupling constant and 
the radius of $S^6$, 
suggesting the stability of $S^6$. 
So the solution obtained in this paper may be 
applied to a compactification such as $AdS_4 \times {\mathbb C} P^3$ 
which may have some relation to string theory. 

This paper is organized as follows. 
In Sec.~\ref{sec1} we give a pseudo-energy 
on general manifolds of even dimensions, 
which realize self-duality relations among higher order terms of the field strength 
as Bogomol'nyi equations.
In Sec.~\ref{sec2} we construct an explicit solution 
in the case that the base manifold is the complex projective space
${\mathbb C}P^n$. 
Sec.~\ref{sec3} is devoted to conclusion and discussion. 
In Appendix we summarize basic ingredients of 
the complex projective space ${\mathbb C}P^n$, 
including Hodge dual relations among 
higher order terms of the curvature tensor  
crucially used in this paper.


\section{Duality as Bogomol'nyi equation}\label{sec1}
We study  ${\rm U}(n)$ 
gauge theory with higher derivative 
couplings in $2n$ dimensions.  
First we consider a pseudo-energy introduced by 
Tchrakian \cite{Tchrakian:1978sf}. 
In this section we work on a general 
$2n$-dimensional Riemannian manifold $M$ with a metric $G$, 
and we will restrict ourselves to the case of 
$M= {\mathbb C} P^n$ when we consider an explicit solution 
in the next section.
On a generic atlas, we parameterize it by the coordinates $x^{\mu}$ ($\mu =1,\cdots, 2n$).
Gauge fields are represented as a $u(n)$-valued one-form $A$ 
which is an $n \times n$ matrix with coefficients of differential forms. 
The field strength two-form $F$ is defined as $F=dA + A \wedge A$.
Let $T^a ~(a=1,2,\cdots ,g:=n^2)$ be generators of the Lie algebra $u(n)$. 
The gauge field $A$ can be expanded by these basis; $A=A^a_m T^a dx^m$. 

We consider ${\rm U}(n)$ 
 gauge systems whose 
pseudo-energy is given by a sum of energy functionals
\begin{align}
 {\mathcal E} &= \sum_{p=0}^{n} \left( \alpha_p^2 {\mathcal E}_{p}^{(1)}  
 + \beta_p^2 {\mathcal E}_p^{(2)}\right) ,
\label{eq:genene}
\end{align}
with $\alpha_p, \beta_p$ real positive constants, 
and energy functionals ${\mathcal E}_{p}^{(1)}$ and ${\mathcal E}_{p}^{(2)}$ defined by
\begin{align}
 {\mathcal E}_{p}^{(1)} &=  \int_{M} {\rm Tr}  {\boldsymbol e}_p(F) \wedge *  {\boldsymbol e}_p(F) ,&
 {\mathcal E}_{p}^{(2)} &= \int_{M}   {\rm ch}_p(F) \wedge *  {\rm ch}_p(F) .
\end{align}
Here ${\boldsymbol e}_p(F)$ and 
the $p$-th Chern character ${\rm ch}_p(F)$
are defined by
\begin{align}
{\boldsymbol e}(F) &:= \exp \left(  \frac{{\bf i} F}{2 \pi} \right) = \sum_{p=0}^{\infty} {\boldsymbol e}_p(F) , &
{\boldsymbol e}_p(F) &:=  \frac{1}{p!} \left(  \frac{{\bf i} F}{2 \pi} \right)^p\cr
{\rm ch}(F) &:= {\rm Tr} \exp \left(  \frac{{\bf i} F}{2 \pi} \right)  = \sum_{p=0}^{\infty} {\rm ch}_p(F) , &
{\rm ch}_p(F) &:=   {\rm Tr} \frac{1}{p!} \left(  \frac{{\bf i} F}{2 \pi} \right)^p = {\rm Tr} {\boldsymbol e}_p(F) ~
 \label{eq:def-e}
\end{align} 
with ${\rm ch}(F)$ the total Chern character. 
${\boldsymbol e}_p(F)$ and ${\rm ch}_p(F)$ are Hermitian, and 
${\rm ch}_p(F)$ is a closed form locally written 
as $ {\rm ch}_p(F) = d \omega_{2p-1}(A)$ with 
the Chern-Simons form
 $\omega_{2p-1}(A)$. 
We also use the following combination
\begin{align}
{\bf K}(F) &= {\boldsymbol e}(F)  - \frac{1}{n} {\rm ch}(F) {\bf 1}_n , & 
{\bf K}_p(F) &= {\boldsymbol e}_p(F)  - \frac{1}{n} {\rm ch}_p(F)   {\bf 1}_n ,
 \label{eq:def-K}
\end{align}
which are traceless parts of ${\boldsymbol e}(F)$ and ${\boldsymbol e}_p(F)$, 
respectively. 
As we will see later, 
our self-duality relation 
defined below 
can be split into the relation between the pieces with respect 
to the decomposition by these basis. 
 
In this paper, we call ${\mathcal E}_{p}^{(1)}$ and ${\mathcal E}_{p}^{(2)}$ 
single and double trace terms, respectively. 
We will not involve triple or higher trace terms.
Although our model is not the most general in this sense, 
we will see that 
the double trace term is sufficient to be considered 
in order to support topological solitons.

We consider Bogomol'nyi equations derived from the pseudo-energy
(\ref{eq:genene}), 
which can be written in the form of a self-duality relation with respect to the Hodge dual operator $*$ induced by the metric as
\begin{align}
  \alpha_p {\bf K}_p(F) &= 
*  \epsilon_1  \alpha_{n-p} {\bf K}_{n-p}(F), 
&\tilde{\beta}_p {\rm ch}_p(F)   &= 
 * \epsilon_2 \tilde{\beta}_{n-p} {\rm ch}_{n-p}(F)   
\label{eq:gen-bps}
\end{align}
where $p=0,1,\cdots,[n/2]$, $\epsilon_1^2=\epsilon_2^2=1$ and 
we have defined 
\begin{align}
 \tilde{\beta}_p^2 :=\beta_p^2+ \alpha_p^2/n.  \label{eq:beta}
\end{align}
Here we have used an equation
\begin{align}
{\rm Tr} {\bf K}_p(F)  \wedge * {\bf K}_p(F)  &={\rm Tr} \left( {\boldsymbol e}_p(F)  - \frac{1}{n} {\rm ch}_p(F)   {\bf 1}_n \right) \wedge * 
\left( {\boldsymbol e}_p(F)  - \frac{1}{n} {\rm ch}_p(F)   {\bf 1}_n \right) \cr
&= {\rm Tr} {\boldsymbol e}_p(F) \wedge * {\boldsymbol e}_p(F) - \frac{1}{n}  {\rm ch}_p(F)  \wedge * {\rm ch}_p(F)  .
\end{align}
In Eq.(\ref{eq:gen-bps}) 
$\epsilon_{1,2}$ denote signs which are deterimend 
in order that 
the Bogomol'nyi bound given below becomes positive. 
Eq.(\ref{eq:gen-bps}) is preserved by
a simultaneous multiplication of a constant $\lambda\, (\neq 0)$ 
on parameters in the pseudo-energy, 
$(\alpha_p,\alpha_{n-p}, \beta_p,\beta_{n-p}) \rightarrow (\alpha_p',\alpha_{n-p}', \beta_p',\beta_{n-p}') = \lambda (\alpha_p,\alpha_{n-p},  \beta_p,\beta_{n-p})$. 
In general, there are $[n/2]$ pairs of equations of this type and in such a generic case it is too hard to discuss about a nontrivial solution. 
Therefore we will restrict our attention to the case in which 
only one pair of equations exists. 
Let us restrict the integer $p$ as $1 \leq p \leq [n/2]$ and 
suppose that nonzero parameters are only $\alpha_p,\alpha_q,\beta_p, \beta_q$ with $q:=n-p$.  
Then our pseudo-energy has only four terms as
\begin{align}
E_{p,q}[\alpha_p,\beta_p,\alpha_q,\beta_q ;F] &=  \alpha_p^2 {\mathcal E}_{p}^{(1)}   +  \alpha_q^2 {\mathcal E}_{q}^{(1)} + \beta_p^2 {\mathcal E}_p^{(2)} + \beta_q^2 {\mathcal E}_q^{(2)} .
 \label{eq:pseudo-four-terms}
\end{align} 
This pseudo-energy can be rewritten in terms of 
${\bf K}_{p,q}(F)$ and ${\rm ch}_{p,q}(F)$ as
\begin{align}
E_{p,q}[\alpha_p,\beta_p,\alpha_q,\beta_q ;F]
 &= \int_{M} \bigg\{   \alpha_p^2 {\rm Tr} {\bf K}_p(F)  \wedge * {\bf K}_p(F)  + \alpha_q^2 {\rm Tr}  {\bf K}_q(F)  \wedge * {\bf K}_q(F) \cr
 & \quad + \tilde{\beta}_p^2 {\rm ch}_p(F) \wedge * {\rm ch}_p(F) +  \tilde{\beta}_q^2 {\rm ch}_q(F) \wedge * {\rm ch}_q(F)  \bigg\}  ,
\end{align}
where coefficients $\tilde{\beta}_{p}$ and $\tilde{\beta}_{q}$ 
are defined in (\ref{eq:beta}). 
This pseudo-energy exhibits the Bogomol'nyi completion 
\begin{align}
E_{p,q}[\alpha_p,\beta_p,\alpha_q,\beta_q ;F]
&= \int_M {\rm Tr}  \left( \alpha_p  {\bf K}_p(F)  - \epsilon_1 * \alpha_q  {\bf K}_q(F)   \right) \wedge  * 
\left( \alpha_p  {\bf K}_p(F)  -* \epsilon_1  \alpha_q  {\bf K}_q(F)   \right)  \cr
 &\quad + \int_M \left( \tilde{\beta}_p {\rm ch}_p(F) -*  \epsilon_2 \tilde{\beta}_q {\rm ch}_q(F)  \right) \wedge * 
\left( \tilde{\beta}_p {\rm ch}_p(F) - * \epsilon_2 \tilde{\beta}_q {\rm ch}_q(F)  \right)  \cr
&\quad + 2  \epsilon_1  \alpha_p \alpha_q \int_M {\rm Tr} {\bf K}_p(F)  \wedge {\bf K}_q(F) 
+ 2 \epsilon_2 \tilde{\beta}_p \tilde{\beta}_q \int_M {\rm ch}_p (F) \wedge   {\rm ch}_q (F) .
\label{eqn:Bogomolnyi1}
\end{align}
Here the integrands in the first two lines 
are positive definite, and therefore the pseudo-energy is bounded from below 
as follows 
\begin{align}
E_{p,q}[\alpha_p,\beta_p,\alpha_q,\beta_q ;F]
& \geq  2 \epsilon_1 \alpha_p \alpha_q  \int_M {\rm ch}_n (F)  + 2  \left( \epsilon_2 \tilde{\beta}_p \tilde{\beta}_q  - \frac{\epsilon_1 \alpha_p \alpha_q}{n}  \right)  \int_M {\rm ch}_p (F) \wedge   {\rm ch}_q (F)  \cr
&=:Q_p[\alpha_p,\beta_p,\alpha_q,\beta_q, \epsilon_1 , \epsilon_2  ; F].
\label{eq:bog1}
\end{align}
The quantity $Q_p[\alpha_p,\beta_p,\alpha_q,\beta_q , \epsilon_1 , \epsilon_2 ; F]$ 
is defined by the integration of an exact form over the whole space and is topological.  
The pseudo-energy is bounded from below by it, 
and the equality is saturated by configurations satisfying 
the Bogomol'nyi equations (\ref{eq:gen-bps}) as expected.

\section{A BPS solution on the projective space} \label{sec2}
In this section we construct 
an explicit solution to the Bogomol'nyi equations (\ref{eq:gen-bps}) 
when the base space is ${\mathbb C} P^{n}$ with the Fubini-Study metric.  
Basics of this space used in this paper are summarized in Appendix. 
We work on a patch whose points are parameterized by an $n$-dimensional complex 
(column) vector $W$. 
The Fubini-Study metric and the corresponding K\"ahler two-form 
are given by
\begin{align}
 ds^2  &= \frac{|dW|^2}{1+r^2} - \frac{|W^{\dag} dW|^2}{(1+r^2)^2}, &
 J&:= \frac{1}{4 {\bf i}} d \left( \frac{W^{\dag} dW - dW^{\dag} W}{(1+r^2)} \right) , \label{eq:Fubini}
\end{align}
respectively, with $r^2 = W^{\dag} W$. 
We show that the following gauge field $A_0$ and the corresponding field strength $F_0$ give a solution to
the Bogomol'nyi equations (\ref{eq:gen-bps}) 
if the parameters in the pseudo-energy satisfy certain relations:
\begin{align}
A_0 &= \frac{1}{r^2}(1-\cos \rho) (W dW^{\dag} - dW W^{\dag} )  + \frac{1}{2r^4} (1-\cos \rho)^2 (W^{\dag} dW - dW^{\dag} W) WW^{\dag} \cr
F_0 &:= d A_0 + A_0 \wedge A_0 = V \wedge V^{\dag}, 
\label{eqn:spinconn}
\end{align}
where  $\cos \rho = 1/\sqrt{1+r^2}$ and  
the one-form valued vector $V$ denotes the corresponding vielbein, 
given as [see Eq.~(\ref{eq:V})]
\begin{align}
V &= \cos \rho dW - \frac{1}{r^2} ( 1- \cos \rho ) \cos \rho W W^{\dag} dW.
\end{align}
The gauge field $A_0$ coincides with a part of the spin connection (\ref{eq:omega-xi}) of the Fubini-Study metric, 
and the field strength $F_0$ coincides with the curvature 2-form $R$ in
Eq.~(\ref{eqn:curvature}).
In the previous section we consider the Hermitian matrix-valued differential form  ${\bf K}_p(F)$ and the Chern character ${\rm ch}_p(F)$ for a field strength $F$. 
Let us substitute $F=F_0$ and consider quantities ${\bf K}_p(F_0)$ and ${\rm ch}_p(F_0)$. 
The field strength $F_0$ satisfies the following duality relations 
\begin{align}
* p \pi^p {\bf K}_p(F_0) &= - (-1)^{n(n+1)/2} q  \pi^{q} {\bf K}_{q}(F_0)\cr
*  \pi^p  {\rm ch}_p(F_0)  &=  (-1)^{n(n+1)/2}   \pi^{q} {\rm ch}_{q}(F_0) 
 \label{eq:bps-2}
\end{align}
as shown in (\ref{eq:relations-R}) in which 
the curvature 2-form $R$ is replaced here by the field strength $F_0$. 
The field strength $F_0$ and the K\"ahler 2-form
$J$ in (\ref{eq:Fubini}) satisfy 
[see Eq.~(\ref{eq:R^kJ^k})]
\begin{align}
{\rm Tr} \wpow{F_0}{k} &= - ( 2 {\bf i} )^{k} \wpow{J}{k}  .
\label{eqn:FpowJ}
\end{align}
Comparing Eq.~(\ref{eq:bps-2}) with Eq.~(\ref{eq:gen-bps}), 
the parameters in the pseudo-energy should be chosen as
\begin{align}
\alpha_p &= \lambda_1 p \pi^p, &
\alpha_q &= \lambda_1 q \pi^q, \cr
\tilde{\beta}_p &= \lambda_2 \pi^{p}, & 
\tilde{\beta}_q &= \lambda_2 \pi^{q}. \label{eq:relations}
\end{align}
When the parameters in the pseudo-energy satisfy these relations 
with certain parameters $\lambda_1, \lambda_2$,
 the connection $A_0$ can be a solution to the Bogomol'nyi equations with the signs
\begin{align}
\epsilon_1 &= -(-1)^{n(n+1)/2}, &
\epsilon_2 &= (-1)^{n(n+1)/2} .\label{eq:signs}
\end{align}
From Eqs.~(\ref{eq:beta}) and (\ref{eq:relations})
the parameters $\beta_p$ and $\beta_q$ are obtained as
\begin{align}
\beta_p &= \sqrt{ \tilde{\beta}_p^2 - \frac{\alpha_p^2}{n} } = \lambda_1 \sqrt{ \mu^2 - \frac{p^2}{n^2} } \pi^p =: \lambda_3(p) \pi^p ,\cr
\beta_q &= \sqrt{ \tilde{\beta}_q^2 - \frac{\alpha_q^2}{n} } = \lambda_1 \sqrt{ \mu^2 - \frac{q^2}{n^2} } \pi^q =: \lambda_3(q) \pi^q 
\end{align}
with $\mu := \lambda_2 / \lambda_1$.
This implies that in the parameter region, $\mu^2 < q^2/n^2$, $\beta_q$ becomes pure imaginary and 
the connection $A_0$ fails to be a solution to the Bogomol'nyi equations.
The pseudo-energy (\ref{eq:pseudo-four-terms}) contains four free parameters $\alpha_p, \alpha_q, \beta_p, \beta_q$. 
The gauge field $A_0$ solves the Bogomol'nyi equations 
if these parameters satisfy the relations $\alpha_p/\alpha_q = p \pi^p / q \pi^q$ and $\beta_p /\beta_q = \pi^p/\pi^q$. 

The topological charge of this configuration 
can be calculated, to give 
\begin{align}
&Q_p[\lambda_1 p \pi^p , 
\lambda_3(p)
 \pi^p, \lambda_1 q \pi^q, 
\lambda_3(q)
 \pi^q, -(-1)^{n(n+1)/2} , (-1)^{n(n+1)/2}  ; F_0] \cr
&=  - (-1)^{n(n+1)/2} 2 \lambda_1^2  \pi^n \left\{   pq  \int_M {\rm ch}_n (F)  -   \left(  \mu^2   + \frac{ pq}{n}  \right)  \int_M {\rm ch}_p (F) \wedge   {\rm ch}_q (F)  \right\} \cr
 &=  - (-1)^{n(n+1)/2} 2 \lambda_1^2  \pi^n \left\{   pq \frac{1}{n!} \int_M  {\rm Tr} \left( \frac{{\bf i} F_0}{2 \pi}   \right)^n  -   \left(  \mu^2   + \frac{ pq}{n}  \right)  \frac{1}{p!q!} \int_M {\rm Tr}\left( \frac{{\bf i} F_0}{2 \pi}   \right)^p
{\rm Tr} \left( \frac{{\bf i} F_0}{2 \pi}   \right)^q  \right\} \cr
&=   (-1)^{n(n-1)/2} 2 \lambda_1^2 
\left\{   pq \frac{1}{n!}   +   \left(  \mu^2   + \frac{ pq}{n}  \right)  \frac{1}{p!q!}   \right\} \int_M \wpow{J}{n}\cr
&=   2 \lambda_1^2 
\left\{   pq \frac{1}{n!}   +   \left(  \mu^2   + \frac{ pq}{n}  \right)  \frac{1}{p!q!}   \right\} n! \int_M dv\cr
&=   2 \lambda_1^2 
\left\{   pq    +   \left(  \mu^2    + \frac{ pq}{n}  \right)  {}_n C_p   \right\} \frac{\pi^n}{n!} .
\end{align}
Here the relation (\ref{eqn:FpowJ}) reduces the integrand in the third line into the power $\wpow{J}{n}$ of the K\"ahler 2-form $J$. 
The resultant is proportional to the volume form $dv$ of  ${\mathbb C} P^n$.

Before closing this section we comment on the relation with self-duality relations on the six-dimensional sphere \cite{Kihara:2007di,Kihara:2007vz}. 
In our previous work \cite{Kihara:2007di,Kihara:2007vz}, a connection, obtained by embedding 
the spin connection $\omega$ of the standard metric on $S^6$ into $8 \times 8$ Hermitian matrices in terms of the Clifford algebra, 
solves similar self-duality equation;  ${\bf K}_1[F_{\omega} ] = \gamma_7 * \epsilon   {\bf K}_2[F_{\omega} ] $. Here 
$F_{\omega}$ is the corresponding field strength, the gauge group is SO(6), and $\gamma_7$ is the chirality matrix 
in the Clifford algebra. 
It gives a twisted closed form ${\rm ch}(F; \gamma_7) := {\rm Tr} \gamma_7 {\boldsymbol e}(F)$. 

In a generalized gauge theory whose gauge group is unitary group ${\rm U}(n)$ 
 we can consider such a twisted operator ${\rm ch}(F; \phi) := {\rm Tr} \phi {\boldsymbol e}(F)$ where $\phi$ is a covariantly constant operator {\it i.e.} $D_A\phi := d \phi + [A , \phi] =0$. The condition ensures that the differential form ${\rm ch}(F; \phi)$ is closed and its integration gives a topologically invariant quantity. 
For fixed $x$, $\phi(x) $ is a Hermitian matrix. We can diagonalize $\phi(x)$ with a unitary matrix $U(x)$; $\varphi(x) = U(x) \phi(x) U(x)^{-1}$.   Therefore the equation becomes $d \varphi + [ \tilde{A}, \varphi ]=0$ where $\tilde{A} = U A U^{-1} + U d U^{-1}$. In addition the  commutator part $[ \tilde{A}, \varphi ]$ has no diagonal element. 
Therefore we conclude that $\varphi$ is locally constant and it commutes with $\tilde{A}$; $[\tilde{A} , \varphi ]=0$. 
If the representation is irreducible as a representation of ${\rm SU}(n)$, 
 such a matrix must be scalar operation and it becomes the type of Eq.~(\ref{eq:gen-bps}). 
In the model on the six-dimensional sphere, which was commented above, 
 $\gamma_7$ is constant and commute with all elements of SO(6) embedded into $8 \times 8$ Hermitian matrices. 
In this case, the matrix $\gamma_7$ is not a scalar matrix
; $\gamma_7 \neq c {\bf 1}$ for any $c \in {\mathbb C}$.
 Actually this representation is reducible as a representation of the SO(6).

\section{Conclusion and Discussion} \label{sec3}
In this paper, 
we have studied the generalized self-duality relations 
of ${\rm U}(n)$ 
 gauge group and the pseudo-energies which lead them. 
After we discussed the equations on general $2n$ dimensional manifolds, 
we have constructed a model on the complex projective space 
${\mathbb C}P^n$, 
the Bogomol'nyi equations on which are 
solved by the spin connection with respect to the Fubini-Study metric. 
The self-duality relations include double trace terms,  
implying that it does not solve the self-duality equations 
previously considered on $S^6$ 
which contain only single trace term 
\cite{Kihara:2007di,Kihara:2007vz}.  
Originally these solutions were considered by Tchrakian {\it et. al.} \cite{Radu:2007az}. 
We have revised it and have found non-vanishing pseudo-energy.

Our solution exists only when the parameters in the pseudo-energy (\ref{eq:pseudo-four-terms}) satisfy 
the relations $\alpha_p/\alpha_q = p \pi^p / q \pi^q$ and $\beta_p /\beta_q = \pi^p/\pi^q$. 
At this stage 
we do not know if another solution exists when these relations are 
not satisfied. 
However the Bogomol'nyi bound exists for any parameter region, 
and so there should exist a solution. 
Our solution satisfies the Bogomol'nyi equations with definite signs 
for $\epsilon_1$ and $\epsilon_2$ given in Eq.~(\ref{eq:signs}).  
We have not yet known if a different sign asignment allows any solution.
Our solution is topological in the sense that energy bound is saturated by a topological quantity, but we have not specified
what kind of homotopy is related to this.
We have constructed a solution with the minimum topological charge {\it i.e.}
a single soliton. 
Multiple solitons and their dynamics deserve to be studied. 
Dimensional reductions to lower dimensions will introduce Higgs scalar fields. 
In this case there will be monopole-like solutions \cite{Kihara:2004yz} too. 

We could consider more general pseudo-energy of the form in Eq.~(\ref{eq:genene}) containing several terms in general more 
than four as considered in this paper. 
Moreover we have not considered multi-trace terms, 
so an extension including them remains as a future problem. 

Finally it is interesting to apply the solution found in this paper 
to a compactification of higher dimension \cite{Kihara:2007vz,Radu:2007az} 
such as $AdS_4 \times {\mathbb C} P^3$ 
which may have a potential use in string theory \cite{CP^3}. 

{\bf Acknowledgments : }
We would like to thank  T. Tchrakian for his valuable comments.  
H.K. is grateful to K. Tsuda and K. P. Yogendran for their comments. 
The work of M.N. is supported in part by Grant-in-Aid for Scientific
Research (No.~20740141) from the Ministry
of Education, Culture, Sports, Science and Technology-Japan.

\appendix

\section{Duality Relations on Complex Projective Spaces}
In this appendix we summarize properties of 
the complex projective space such as 
higher order wedge products of the curvature 2-form and 
their Hodge dual, 
which will be used in 
this paper. 
The {\it $n$-dimensional complex projective space} ${\mathbb C} P^n$ 
is defined as a family of one-dimensional subspaces in an $(n+1)$-dimensional vector space ${\mathbb C}^{n+1}$.
It can be written as a quotient space $( {\mathbb C}^{n+1} \setminus \{ 0 \} )/{\mathbb C}^{\times}$, where ${\mathbb C}^{\times} = {\mathbb C} \setminus \{ 0 \}$. {\it Homogeneous coordinates} of ${\mathbb C}^{n+1}$
are used as representatives; $[Z_1 : \cdots : Z_{n+1} ]$. 
An equivalence relation is introduced by 
scalar multiplications: $[\alpha Z_1 : \cdots : \alpha Z_{n+1}]= [Z_1 : \cdots : Z_{n+1} ]$,~($\alpha \neq 0$). 
The space ${\mathbb C} P^n$ is covered by $(n+1)$-patches $U_i$; $U_i := \{[Z_1 : \cdots : Z_{n+1} ] \in {\mathbb C} P^n | Z_i \neq 0  \}$. 
Each patch $U_i$ is isomorphic to ${\mathbb C}^n$  
where the isomorphism is given by 
$W_j^{(i)} = Z_j/Z_{i}$,~$(i \neq j)$. 
 On the patch $U_{n+1}$ we can choose a local coordinate system $W=(W_1 , \cdots ,W_n)^t \in {\mathbb C}^n$
 ($W_j:=W_j^{(n+1)}$). 
Suppose that the space is endowed with the {\it Fubini-Study metric}, 
which is given as a line element on the patch $U_{n+1}$;
\begin{align}
ds^2 &= \frac{(1+|W|^2) |dW|^2  - |W^{\dag} dW|^2}{(1+|W|^2)^2}~~,& |W|^2 &= W^{\dag} W~~.
\end{align}
Because of the symmetry, 
this can be rewritten in the same form in other patches. 
This space is not only a Riemannian but also an Hermitian manifold. 
It can be realized 
as a homogeneous space ${\mathbb C} P^n = {\rm U}(n+1)/({\rm U}(n)\times {\rm U}(1))$. 
The Fubini-Study metric can be identified with 
a part of the Maurer-Cartan form as seen later.
We define the radial quantity $r = |W|$ and a function $\rho(r) := \arctan r$.   
The corresponding Lie algebra $u(n+1)$ is a set consisting 
of anti-Hermitian matrices. 
The algebra $u(n+1)$ split into two parts; $u(n) \oplus u(1)$ and $cp(n)$. 
\begin{align}
u(n) \oplus u(1) &:= \{ {\bf i} X \in u(n+1) | [\phi , X]=0  \}~,\cr
cp(n)  &:= \{ {\bf i} X \in u(n+1) | \{\phi , X\}=0  \}~,\cr
~~ \phi &
:=
 \begin{pmatrix}
{\bf 1}_n&\\
&-1
\end{pmatrix}~~,
\end{align}
where ${\bf 1}_n$ is the $n \times n$ unit matrix and the matrix $X$ is traceless and Hermitian. 
Note that each element in the subspace $u(n) \oplus u(1)$ is block-diagonal $A \oplus B$ where $A$ is an $n \times n$ matrix and $B$ is a complex number. 
 The matrix $\varphi= {\rm diag}({\bf 1}_{n}, -n)$
\footnote{The matrix $\varphi$ differs from $\phi$.}
 is a generator of the $u(1)$ part. 
A generator in $cp(n)$ is represented by a matrix with the following form:
\begin{align}
X &= 
\begin{pmatrix}
& W\\
W^{\dag}&
\end{pmatrix}~~.
\end{align}
We often use the normalized matrix $\hat{X} = \frac{1}{|W|} X$ 
for convenience. 
The exponential of this generator becomes an element in group ${\rm U}(n+1)$.  
\begin{align}
 g = \exp({\bf i} \rho \hat{X}) &= \sum_{k=0}^{\infty} \frac{1}{k!} ({\bf i} \rho \hat{X})^{k} 
= {\bf 1}_{n+1} + ( \cos \rho -1  ) \hat{X}^2+ {\bf i} \sin \rho  \hat{X} ~~.
\end{align}
where we have inserted a scale factor $\rho$ 
which we leave arbitrary in order to obtain the Fubini-Study metric.
Here we have used
\begin{align}
\hat{X}^2  &=  
\begin{pmatrix} \hat{W} \hat{W}^{\dag}  & \\ & 1 \end{pmatrix} ~, 
 \quad
\hat{X}^3 
=\hat{X}~.
\end{align}
The inverse of $g$ is its Hermitian conjugate $g^{\dag}$: $g^{\dag}g =1$. 
The {\it Maurer-Cartan form} $\alpha_{_{\rm MC}}$ is defined as
\begin{align}
\alpha_{_{\rm MC}} = g^{-1} dg&= (1-\cos \rho ) ( \hat{X} d \hat{X} - d \hat{X} \hat{X} )  + (\cos \rho -1 )^2 \hat{X}^2 d \hat{X} \hat{X}\cr
&+{\bf i} \left( \hat{X} d\rho + \sin \rho d \hat{X} +(1-\cos \rho )\sin \rho \hat{X} d \hat{X} \hat{X}   \right)~~.
\end{align}
This takes a value 
in the Lie algebra $u(n+1)$ splitting into two parts with respect to the decomposition $u(n+1) = (u(n)\oplus u(1)) \oplus cp(n)$: 
$\alpha_{_{\rm MC}} = \tilde{\omega} + {\bf i} E$ with 
\begin{align}
\tilde{\omega} &=  (1-\cos \rho ) ( \hat{X} d \hat{X} - d \hat{X} \hat{X} ) 
 + (\cos \rho -1 )^2 \hat{X}^2 d \hat{X} \hat{X} \in u(n) \oplus u(1), ~~ \cr
E &= \hat{X} d\rho + \sin \rho d \hat{X} +(1-\cos \rho )\sin \rho \hat{X} d \hat{X} \hat{X}
 \in cp(n) ~~.
\end{align}
The matrix valued one-form $E$ gives the vielbein $V$ 
with respect to the Fubini-Study metric 
whereas the $u(n) \oplus u(1)$ part gives the corresponding spin connection.  
First the vielbein is obtained from $E$ by  
\begin{align}
 E =
\begin{pmatrix}
           0_n & V \\
     V^\dagger & 0_n
\end{pmatrix} 
\end{align}
to yield 
\begin{align}
V &= \cos \rho dW - \frac{1}{r^2} ( 1- \cos \rho ) \cos \rho W W^{\dag} dW~~,& \cos \rho &= \frac{1}{\sqrt{1+r^2}}~~. \label{eq:V}
\end{align}
The Fubini-Study metric can be obtained by  
the inner product of $V^{\dag}$ and $V$ with the symmetric product as 
\begin{align}
 ds^2 = V^{\dag} \cdot V = \sum_i V_i^* V_i
 &= \frac{|dW|^2}{1+r^2} - \frac{|W^{\dag} dW|^2}{(1+r^2)^2}~~, 
\end{align}
and 
the wedge product of them gives 
the K\"ahler two-form 
\begin{align}
J & :=  \frac{1}{2 {\bf i}} V^{\dag} \wedge V= \frac{1}{2 {\bf i}} \left( \frac{dW^{\dag} \wedge dW}{1+r^2} - \frac{dW^{\dag} W W^{\dag} \wedge dW}{(1+r^2)^2} \right)~~.
\end{align}
This form is closed and at least on the patch $U_{n+1}$ it can be written as an exact form:
\begin{align}
J &= \frac{1}{4 {\bf i}} d \left( \frac{W^{\dag} dW - dW^{\dag} W}{(1+r^2)} \right)~~,
\label{eqn:potential}
\end{align}
where this equation does not mean that the form $J$ is exact all around the space ${\mathbb C} P^n$. 
The form $J$ is a nontrivial closed form, so the equivalence class including the form $J$ is a non-zero element of the cohomology ring $H^*({\mathbb C} P^n)={\rm Ker}(d : \Lambda^*({\mathbb C} P^n) \rightarrow \Lambda^*({\mathbb C} P^n) ) / {\rm Im}(d : \Lambda^*({\mathbb C} P^n) \rightarrow  \Lambda^*({\mathbb C} P^n))$. 
Second, we consider the $u(n) \oplus u(1)$ part $\tilde{\omega}$, which is a spin connection with respect to the Fubini-Study metric. 
As mentioned above, generators in the $u(n) \oplus u(1)$ part are block-diagonal. We separate them into an $n \times n$ matrix-valued one-form $\omega$ and a one-form $\xi$ with pure imaginary coefficient, 
\begin{align}
\tilde \omega &
 = \begin{pmatrix}
    \omega & 0 \\ 
         0 & \xi
   \end{pmatrix}, \\
\omega &= \frac{1}{r^2}(1-\cos \rho) (W dW^{\dag} - dW W^{\dag} )  + \frac{1}{2r^4} (1-\cos \rho)^2 (W^{\dag} dW - dW^{\dag} W) WW^{\dag}~~,\cr
\xi &= \frac{1}{2} \cos^2 \rho (W^{\dag} dW - dW^{\dag} W)~~.
\label{eq:omega-xi}
\end{align}
Here the form $\xi$ is a gauge potential of the K\"ahler two-form $J$ as shown in (\ref{eqn:potential}).
They form actually a spin connection: 
\begin{align}
dV + \omega \wedge V + V \wedge \xi &=0~~. 
\end{align} 
The curvature 2-form $R$ 
obtained from the $n \times n$ block $\omega$ is computed as
\begin{align}
R &:= d \omega + \omega \wedge \omega = V \wedge V^{\dag}~~,&
 {\rm Tr} R &= - V^{\dag} \wedge V = - (2 {\bf i}) J~~.
\label{eqn:curvature}
\end{align}

Next we calculate powers of the curvature $R$ and their Hodge dual. 
The above relation (\ref{eqn:curvature}) shows that the exponents of the curvature $R$ are written as
\begin{align}
\wpow{R}{s} &= (2{\bf i})^{s-1} \wpow{J}{s-1} \wedge R~~.
\end{align}
In order to calculate its Hodge dual, 
we define the Hodge dual on forms. 
Let us decompose the vielbein $V$ in (\ref{eq:V}) 
in the real space;  
\begin{align}
V_i &= \sigma_i + {\bf i} \tau_i~~,& ds^2 & 
 = \sum_i \{ (\sigma_i)^2 + (\tau_i)^2 \}~~.
\end{align}
Here $\sigma_i$ and $\tau_i$ are differential forms with real coefficients, that is to say, $\sigma_i^{\dag}  = \sigma_i$, $\tau_i^{\dag}  = \tau_i$.   
They are vielbeins in the real space. 
The invariant volume form with respect to the Fubini-Study metric 
can be written by the real vielbeins as
\begin{align}
dv &:= \sigma_{1 \cdots n} \wedge \tau_{1 \cdots n} 
 = \sigma_1 \wedge \sigma_2 \cdots \wedge \sigma_n \wedge 
   \tau_1 \wedge \tau_2 \cdots \wedge \tau_n
 ~~ ,
\end{align}
integration of which gives the volume.  
The metric is expressed as a matrix $G$ and 
its determinant is denoted by $g = \det G$. Then we have
\begin{align}
\beta &= \frac{1}{1+r^2}~,& G&=\beta {\bf 1}_{n} - \beta^2 WW^{\dag}~,\cr
G W &= \beta W - \beta^2 r^2 W = \beta^2 W~,& G U &= \beta U~,(W^{\dag}U=0).
\end{align}
The eigenvalues of $G$ are $n-1$ $\beta$s and $\beta^2$.  
The determinant $g$ is 
$g= \beta^{2(n+1)}$, and the volume is 
\begin{align}
 {\rm vol}({\mathbb C} P^n) = \int dv &= \frac{\pi^n}{n!}.
\end{align}
Let us introduce a new notation to express $\sigma$ and $\tau$ 
in the same form:
${\mathfrak e}_i := \sigma_i ,~~{\mathfrak e}_{n+i}:=\tau_{i},$~~$1 \leq I_k \leq 2n,~~1 \leq k \leq s$~,
${\mathfrak e}_{I_1 I_2 \cdots I_s} :=  {\mathfrak e}_{I_1} \wedge {\mathfrak e}_{I_2} \wedge \cdots \wedge {\mathfrak e}_{I_s}$.
The Hodge dual operator $*$ on a basis of $s$-forms is defined as
\begin{align}
* {\mathfrak e}_{I_1 \cdots I_s} &= \frac{1}{(2n-s)!} \epsilon_{I_1 \cdots I_s I_{s+1} \cdots I_{2n}} {\mathfrak e}_{I_{s+1} \cdots I_{2n} }~~.
\end{align}
We are ready to compute 
the Hodge duals of $J$, $R$ and their powers.  
We rewrite $J$ and $R$ by $\sigma$ and $\tau$.   
The K\"ahler two-form $J$ and the coefficients of the $k$th-powers of the curvature $R$ are written as 
\begin{align}
V_i^{\dag} \wedge V_i
&= 2 {\bf i} \sigma_i \wedge \tau_i ~~,~~~~
 J = \sum_{i=1}^n \sigma_i \wedge \tau_i~~,\cr
\left( \wpow{R}{k} \right)_{ij} &= (2{\bf i})^{k-1} \wpow{J}{k-1} \wedge \left(   \sigma_i \wedge \sigma_j +\tau_i \wedge \tau_j
- {\bf i} \sigma_i \wedge \tau_j + {\bf i} \tau_i \wedge \sigma_j  \right)~.
\end{align}
The Hodge duals of powers of the K\"ahler two-form $J$ and components of powers of the curvature are summarized: 
\begin{align}
* \wpow{J}{k} &= (-1)^{n(n-1)/2} \frac{k!}{(n-k)!} \wpow{J}{n-k}~~,\cr
* \left( \wpow{J}{k} \wedge \sigma_{ij}  \right) &= -(-1)^{n(n-1)/2} \frac{k!}{(n-k-2)!} \wpow{J}{n-k-2} \wedge \tau_{ij} ,\cr
* \left( \wpow{J}{k} \wedge \tau_{ij}  \right) &= -(-1)^{n(n-1)/2} \frac{k!}{(n-k-2)!} \wpow{J}{n-k-2} \wedge \sigma_{ij} ,\cr
* \left( \wpow{J}{k} \wedge \sigma_ i \wedge \tau_j  \right) &= -
(-1)^{n(n-1)/2} \left\{ \frac{k!}{(n-k-2)!}   \times \wpow{J}{n-k-2} \wedge   ( \sigma_j \wedge  \tau_i  )  \right.\cr
&\left. ~~~~~~~~~~~~~~~~~~~~~~~ - \delta_{ij} \frac{k!}{(n-k-1)!}  \wpow{J}{n-k-1}  \right\}  .
\end{align}
The Hodge duals of the coefficients of $\wpow{R}{k}$ satisfy the following relations:
\begin{align}
*  \left( \wpow{R}{k} \right)_{ij} 
&= - (-1)^{n(n-1)/2} \frac{(k-1)!}{(n-k-1)!} \frac{(2{\bf i})^{k-1}}{(2{\bf i})^{n-k-1}} \left( \wpow{R}{n-k} \right)_{ij}  \cr
&~~~~~ -  (-1)^{n(n-1)/2} \delta_{ij} (2{\bf i})^{k} \frac{(k-1)!}{(n-k)!}\wpow{J}{n-k} \nonumber \\ 
* {\rm Tr} \wpow{R}{k} &= - (-1)^{n(n-1)/2} \frac{(k-1)!}{(n-k-1)!} \frac{(2{\bf i})^{k-1}}{(2{\bf i})^{n-k-1}} {\rm Tr}  \wpow{R}{n-k} \cr
&~~~~~ -  (-1)^{n(n-1)/2} n (2{\bf i})^{k} \frac{(k-1)!}{(n-k)!}\wpow{J}{n-k} 
 \label{eqn:selfdualpowR}
\end{align}
Since the $k$-th exponent of $J$ is proportional to 
${\rm Tr} \wpow{R}{k}$ we have 
\begin{align}
{\rm Tr} \wpow{R}{k} &= - ( 2 {\bf i} )^{k} \wpow{J}{k} ~~.
 \label{eq:R^kJ^k}
\end{align}
Finally, by using notations defined in Eqs.(\ref{eq:def-e}) and 
(\ref{eq:def-K}), 
the relations in Eq.(\ref{eqn:selfdualpowR}) are summarized as
\begin{align}
\epsilon &:= (-1)^{n(n+1)/2}\cr
* k \pi^k {\bf K}_k(R) &= - \epsilon (n-k)  \pi^{n-k} {\bf K}_{n-k}(R)\cr
*  \pi^k  {\rm ch}_k(R)  &=  \epsilon   \pi^{n-k} {\rm ch}_{n-k}(R)  .
 \label{eq:relations-R}
\end{align}
where field strength $F$ in Eqs.(\ref{eq:def-e}) and (\ref{eq:def-K}) 
are replaced by the curvature 2-form $R$.


\end{document}